# $\Omega_{\text{baryon}}$ and the Geometry of Intermediate Redshift Lyman $\alpha$ Absorption Systems

Michael Rauch[1] & Martin G. Haehnelt[2]

1: Carnegie Observatories, 813 Santa Barbara Street, Pasadena, CA 91101, USA

2: Max-Planck Institut für Astrophysik, Karl-Schwarzschild-Straße 1, 85748 Garching, Germany

## Abstract

Estimates of $\Omega_{\text{baryon}}$ from primordial nucleosynthesis together with standard assumptions about the ionization state of low column density Ly$\alpha$ forest clouds can be used to determine an upper limit for the cloud thickness along the line of sight. This upper limit provides significant constraints on the axis ratio of the absorbers and on the overall fraction of the baryonic matter contained in these objects if the recently measured large values for their transversal size are representative. The absorbers have to be considerably flattened structures and may possibly contain a substantial fraction of the baryonic matter at high redshift.

*Subject headings:* Quasars: Absorption Lines — Large Scale Structure of the Universe

Recently it has been emphasized (Petitjean et al. 1993, Meiksin & Madau 1993, Shapiro, Giroux and Babul 1994) that low column density Ly$\alpha$ absorption systems ($10^{13}$cm$^{-2}$ $\leq N \leq$ $10^{15}$cm$^{-2}$) may contain a substantial fraction of the baryonic $\Omega$, if the clouds are highly ionized.

Aside from the intrinsically interesting question of whether the baryons reside predominantly in the intergalactic medium (including Ly$\alpha$ clouds) or in galaxies, the maximum amount of baryonic matter that can be accommodated in Ly$\alpha$ clouds may provide useful constraints on the geometry of these objects (Rauch & Haehnelt 1994).

The baryonic fraction of the closure density in the clouds at intermediate and high redshift redshifts can be computed from the neutral hydrogen (HI) column density distribution function (CDDF), f(N), as described e.g. by Wolfe (1993),

$$\Omega_{Ly\alpha} = \frac{\mu m_H H_0}{c\rho_{0crit}} \int_{N_1}^{N_2} x^{-1}(N) N f(N) dN, \qquad (1)$$

where N and x are the neutral hydrogen column density and the ratio of neutral to total hydrogen, respectively. $H_0$ and $\rho_{0crit}$ are the Hubble constant and the critical density at the present epoch. $\mu m_H$ is the mean molecular weight. We adopt the usual power-law fit to the lower column density range (between $N_1=10^{13}$ and $N_2=10^{15}$cm$^{-2}$), f(N)=$BN^{-\beta}$, with B=$4.9\times 10^7$ and $\beta$=1.46, as given by Hu et al. (1995) for $q_0 = 0$, and integrate between column densities $N_1$ and $N_2$. Assuming that the neutral fraction $x$ of hydrogen is determined by photoionization equilibrium, we have

$$x = 3.9 \times 10^{-6} \left(\frac{T}{3\times 10^4}\right)^{-0.35} \left(\frac{I}{10^{-21}}\right)^{-0.5} \left(\frac{N}{10^{14}}\right)^{0.5} \left(\frac{D}{100\text{kpc}}\right)^{-0.5}, \qquad (2)$$

where $T$ is the gas temperature in K, $I$ is the intensity of the ionizing UV background in units of ergs Hz$^{-1}$ sr$^{-1}$ s$^{-1}$ cm$^{-2}$ (for the adopted fiducial value see Lu, Wolfe, & Turnshek 1991, and refs. therein), $N$ the HI column density in cm$^{-2}$, and $D$ the thickness of the cloud (or the path length of our line of sight through it).

To proceed further we have to make some assumptions about the effective thickness $D$. It may depend on the HI column density $N$ of the cloud but without detailed modeling we can only speculate how. To bracket the likely range of dependence we represent $D$ as a power law in $N$. Here we consider three basic cases:

$$(A) \qquad D = D_0 = \text{const}, \qquad (3)$$

$$(B) \qquad D = D_0 \left(\frac{N}{10^{14}}\right), \text{ and} \qquad (4)$$

$$(C) \qquad D = D_0 \left(\frac{N}{10^{14}}\right)^{-1}, \qquad (5)$$

where $D_0$ is a fiducial thickness which we will express below in units of 100 kpc.

Using case (A) in equation (1) we obtain for the baryonic $\Omega$:

$$\Omega_{\text{baryon}} \approx 0.036 h_{75}^{-1} f_{Ly\alpha}^{-1} \left(\frac{T}{3 \times 10^4}\right)^{0.35} \left(\frac{I}{10^{-21}}\right)^{0.5} \left(\frac{D_0}{100\text{kpc}}\right)^{0.5}. \qquad (6)$$

Here $f_{Ly\alpha}$ is the fraction of baryons contained in Ly$\alpha$ absorption systems. Cases (B) and (C) give $\Omega_{\text{baryon}} \approx 0.047$ and 0.043, respectively, with the same scaling in $T$, $I$, and $D_0$. These $\Omega$ values have to be compared to the range permitted by primordial nucleosynthesis arguments,

$$0.018 \leq \Omega_{nuc} h_{75}^2 \leq 0.027 \qquad (7)$$

(Walker et al. 1991). The nucleosynthesis upper limit on $\Omega$ implies then an upper limit on the size of the absorber along the line of sight,

$$D \leq 56 h_{75}^{-2} f_{Ly\alpha}^2 \left(\frac{T}{3 \times 10^4}\right)^{-0.7} \left(\frac{I}{10^{-21}}\right)^{-1} \text{ kpc}. \qquad (8)$$

Together with the transversal sizes derived from absorption towards multiple QSO images we can derive from this an *upper limit on the typical axis ratio D/L (the ratio of thickness D to transversal length L) of the clouds*: $\Omega_{nuc} \leq 0.027 h_{75}^{-2}$ implies for the most conservative case (A):

$$D/L \leq 0.056 h_{75}^{-2} f_{Ly\alpha}^2 \left(\frac{T}{3 \times 10^4}\right)^{-0.7} \left(\frac{I}{10^{-21}}\right)^{-1} \left(\frac{L}{1 h_{75}^{-1}\text{Mpc}}\right)^{-1}. \qquad (9)$$

Two recent measurements indicate transversal "cloud diameters" (or coherence lengths) of order 1.2 $h_{75}^{-1}$ Mpc for absorbers in the redshift range $0.5 \leq z \leq 0.9$ (Dinshaw et al., 1995), and 360 $h_{75}^{-1}$ kpc or larger at redshift $\sim 1.8$ (Bechtold et al., 1994, Dinshaw et al. 1994). Here $q_0=0$ was assumed.

These values suggest that low column density Ly$\alpha$ absorption systems belong to *significantly flattened* structures, even if the fraction of all baryons contained in these systems were close to one (*c.f.* Barcons & Fabian (1987) who proposed flattened absorbers for quite different reasons).

If this is correct the actual size estimates should be even larger because the values quoted above were derived assuming spherical objects. Flattened structures seen at random aspect angles would have to be larger to subtend the same absorption cross section, so the estimated diameters for $q_0 = 0$ become 1.7 $h_{75}^{-1}$ Mpc ($0.5 \leq z \leq 0.9$) and 500 $h_{75}^{-1}$ kpc ($z \sim 1.8$), respectively.

How are our conclusions affected if transversal cloud sizes change with time ? The size estimates for $z \sim 1.8$ (Bechtold et al. 1994) and $z \sim 0.8$ (Dinshaw et al. 1995) differ at the 95% level, so we have reason to believe that a typical absorber is getting larger with time. Possible explanations include an expansion of the individual absorber (at least along the transverse direction) or formation of absorbers with increasing transversal sizes towards lower redshifts, as expected in a hierarchical structure formation scenario. Comparing the two different size estimates implies an evolution of transversal size roughly according to $(1+z)^{-2.7}$. Extrapolating to a typical absorber redshift in the sample used to determine the CDDF ($z \sim 2.8$) gives a size of $L \sim 200$ kpc for disc-like absorbers. Even when adopting this unfavourable case together with the extreme assumption that all baryons are contained in Ly$\alpha$ absorption systems we find that the axis ratio $D/L$ is required to be less than 0.25.

Note that so far we have only assumed ionization equilibrium, but not thermal equilibrium. Introducing the additional constraint of thermal equilibrium would allow us to use the measured width of the absorption lines to obtain a second independent lower limit on the characteristic density and thus an upper limit on the characteristic thickness and axis ratio of the clouds (see e.g. Donahue & Shull 1991, Hu et al. 1995). However, at present it is not at all clear if heating by photoionization leads to thermal equilibrium. This uncertainty is mainly due to the lack of even a rough estimate of the characteristic density of the clouds. If the density of the clouds is only moderately enhanced compared to the mean baryonic density as suggested in currently favoured models for the Ly$\alpha$ absorption systems (Cen et al. 1994), timescales for photoionization heating and recombination and line cooling are longer than the Hubble time. The temperature of the clouds will then not be determined by photoionization equilibrium, but it will depend on dynamical processes like adiabatic cooling/heating, the spectrum responsible for the reionization of the IGM and the time elapsed since the gas was reionized (Miralda-Escudé & Rees 1994). However, the assumption of total (thermal and ionization) equilibrium would only strengthen our point: with the fiducial values for $T$, $I$ and $N$ adopted here and assumptions for the heating and cooling processes as in Donahue & Shull (1995) the equilibrium thickness of the cloud would be only of order 100 $pc$ (Hu et al. 1995). But one should note here that the exact value will depend very sensitively on the details of the assumptions for the relevant cooling and heating processes and the spectrum of the ionizing background. Work by Cowie et al. (1995) and Hu et al. (1995) based on limits on the relative strength of CIV, CII, SiII and NV (as indicators of the ionization state) suggests that the ionized fraction of the gas may indeed be compatible with a thermal equilibrium model, and we may take this as independent evidence in favour of flattened absorbers.

Can we escape the conclusion that Ly$\alpha$ clouds are seriously flattened ? If our assumptions about the values of the ionizing background radiation, the nucleosynthesis $\Omega_{\text{baryon}}$, and the recent estimates of the transversal size are correct, we can only think of clumping as a potential way to increase the neutral fraction while keeping the overall geometry approximately spherical. Then the individual clumps or cloudlets would have to be confined, presumably in the hot external medium of a (galactic) halo which itself is being kept in place by gravity, a model investigated most recently for low redshift absorption systems by Mo (1994). However, the cloudlets would have to be numerous and of low mass to explain the observed covering factor of order unity. It seems difficult to understand how such cloudlets should form given the rather high Jeans mass in such an environment. Furthermore, the velocity structure seen in high resolution Ly$\alpha$ forest studies indicates multiple component structure only on a velocity scale of a few tens of km/s (e.g. Cowie et al. 1995, Dinshaw et al. 1994), less than expected for halos with galactic or even larger dimensions. Another potential difficulty of the halo picture concerns the confinement itself: assuming a virialized structure with virial velocities of a few tens of km/s the confining halo gas would hardly be hotter than the cloudlets to be confined. Thus small absorbing gas clumps in a quasi-spherical (galactic) halo are unlikely to be an explanation for the large observed absorber sizes at non-zero redshift.

## Conclusions

We have used the assumption that intermediate redshift Ly$\alpha$ forest absorbers are in photoionization equilibrium with the UV background radiation together with limits on $\Omega_{\text{baryon}}$ derived from primordial nucleosynthesis and recent detections of common absorption on large scales to show that these absorbers are likely to be *significantly flattened* structures, with axis ratios of possibly less than 1/10. The recent large size estimates can be reconciled with highly ionized gas in photo-ionization equilibrium if they measure mostly the transversal extent of flat gaseous structures with a large covering factor. The picture of Ly$\alpha$ forest absorbers as flattened structures is well consistent with the conclusions of Cen et al. (1994) and Petitjean, Mücket, and Kates (1995), who find that simulations of gravitational collapse in the universe create filamentary or sheet-like structures giving rise to absorption phenomena very similar to the Ly$\alpha$ forest. It also ties in nicely with the results obtained by Bechtold et al. (1994) and Dinshaw et al. (1994,1995) who find rather small velocity differences among the common absorption line pairs in their double QSO spectra. In fact, the large transversal size-estimates and our result, if correct, do not uniquely distinguish between the existing models (pressure confinement, CDM gravity confinement) suggested earlier for intergalactic clouds; they suggest, however, that most Ly$\alpha$ forest systems, at least at high redshift, do not arise in quasi-spherical structures.

---